\documentclass{IEEE_lsens}

\usepackage{textcomp}
\usepackage{graphicx}

\usepackage[noadjust]{cite}
\ifCLASSINFOpdf
\else
\fi

\usepackage[T1]{fontenc} 
\usepackage{amsmath}
\usepackage[cmintegrals]{newtxmath}
\usepackage{bm}
\usepackage{array}

\usepackage{url}

\ifCLASSINFOpdf
\else
\fi
\providecommand{\hypersetup}[1]{\relax}

\hypersetup{pdftitle={Bare Demo of IEEE\_lsens.cls for IEEE Sensors Letters},
pdfsubject={Typesetting},
pdfauthor={Michael D. Shell},
pdfkeywords={Class, IEEE, IEEE\_lsens, IEEE Sensors Letters, LaTeX, Typesetting, TeX}}

\begin{document}

\markboth{Vol.~0, No.~0, SEPT~2023}{0000000}

\IEEELSENSarticlesubject{Sensor Phenomena}

\title{Frequency-dependent amplitude correction to free-precession scalar magnetometers}

%
\author{\IEEEauthorblockN{Mark E. Limes\IEEEauthorrefmark{1}\IEEEauthorieeemembermark{1}, Lucia Rathbun\IEEEauthorrefmark{2},
Elizabeth Foley\IEEEauthorrefmark{2},Tom~Kornack\IEEEauthorrefmark{2}, Z. Hainsel\IEEEauthorrefmark{3}, and~Alan~Braun\IEEEauthorrefmark{3}}
\IEEEauthorblockA{\IEEEauthorrefmark{1}Virginia Tech, Blacksburg, VA, 24061, USA\\
\IEEEauthorrefmark{2}Twinleaf LLC, Plainsboro, NJ, 08536, USA\\
\IEEEauthorrefmark{3}SRI International, 201 Washington Road,  Princeton, New Jersey, 08540, USA\\
\IEEEauthorieeemembermark{1}Member, IEEE}
\thanks{Corresponding author: M.~E.~Limes (e-mail: mlimes@vt.edu).}\protect
\thanks{Associate Editor: .}%
\thanks{Digital Object Identifier 10.1109/LSENS.2023.0000000}}
%

\IEEELSENSmanuscriptreceived{Manuscript received Oct 30, 2024.
Date of publication 0 , 2024; date of current version0 , 2024.}

\IEEEtitleabstractindextext{%
\begin{abstract}
Pump and probe scalar atomic magnetometers show incredible potential for real-world, traditionally difficult measurement environments due to their high dynamic range and linearity. Previously, it has been assumed these scalar magnetometer have a flat response across their bandwidth, and flat noise floor. Here we show that standard fitting routines, used to extract the magnetic field, result in a non-linear frequency dependent response across the sensor bandwidth, due to the time-averaged nature of such free precession measurements. We present an analytic correction dependent on dead-time, and show how this equation can also correct the sensor spectral density. The maximum in-band amplitude loss approaches 29\% as the frequency of interest becomes the Nyquist frequency, making a significant correction for applications such as source localization in magnetoencephalography. These pump and probe atomic magnetometers also are known to have large aliasing of out-of-band signals, and we propose a scheme where the frequency of out-of-band signals can be identified by performing fits with varying dead-time on the raw free-precession sensor data. 
\end{abstract}

}

\IEEEpubid{1949-307X \copyright\ 2024 IEEE. Personal use is permitted, but republication/redistribution requires IEEE permission.}

\maketitle

The research and commercial development of minature high-performance atomic sensors has matured over the past few decades \cite{Liew_2004,Knappe_2006,Kitching_2018}. Much of this development occurred as a result of key miniaturization innovations, such as single-mode VCSELs and anodically bonded alkali vapor cell technology. 
As an extension of laboratory work done in the early 2000's \cite{Xia_2006}, commercial near-zero field atomic sensors have shown great promise for magnetically shielded magnetoencephalography (MEG) studies that involve minor motion \cite{Boto_2018,Aslam_2023}.
Conventional commercial MEG systems are  cryogenically cooled SQUID arrays that consist of hundreds of gradiometers, that are used within a magnetically shielded room in a clinical setting. Though SQUID systems are currently the dominant commercial technology, there is interest in developing a small, portable MEG system that works unshielded in ambient environments.  
Recently, pump and probe scalar atomic magnetometers that work unshielded in ambient environments have been shown to perform well enough to detect signals from the human brain \cite{Limes_2020}. These scalar magnetometers are useful for such applications, as they a high dynamic range and linearity that stem from a non-calibrated frequency measurement of magnetic field, rather than a calibrated voltage measurement of magnetic field \cite{Li_2011, Sheng_2013, Grujic_2015, Hunter_2018, Gerginov_2020, Hunter_2023}. Currently arrays of the pump and probe scalar magnetometers are being fabricated for use in source localization in MEG trials, where existing MEG algorithms need modified for use with scalar sensors \cite{Clancy_2021}.
 
The bandwidth of pulsed free-precession scalar magnetometers are typically defined by half the repetition rate of consecutive pump and probe cycles, where one magnetic field is extracted during each probe cycle. The resulting response across the sensor bandwidth was previously considered flat. Here, we show a frequency-dependent correction is needed for the response, and give a formula that is able to represent the curvature of the noise floor in the spectral density. Such a correction becomes important for a variety practical use cases of the sensor, including source localization for MEG, when using a sensor where the frequencies of interest are more than half the Nyquist frequency. In addition, free-precession magnetometers based on the magnetic resonance of polarized protons \cite{Waters_1958} or $^3$He \cite{Slocum_1974, Gemmel_2010} have been available commercially and used for a number of years, for magnetic surveying and space-borne magnetometry, and the corrections herein apply to those systems as well.

\IEEEpubidadjcol

\begin{figure}
 	\includegraphics[width=3.0in]{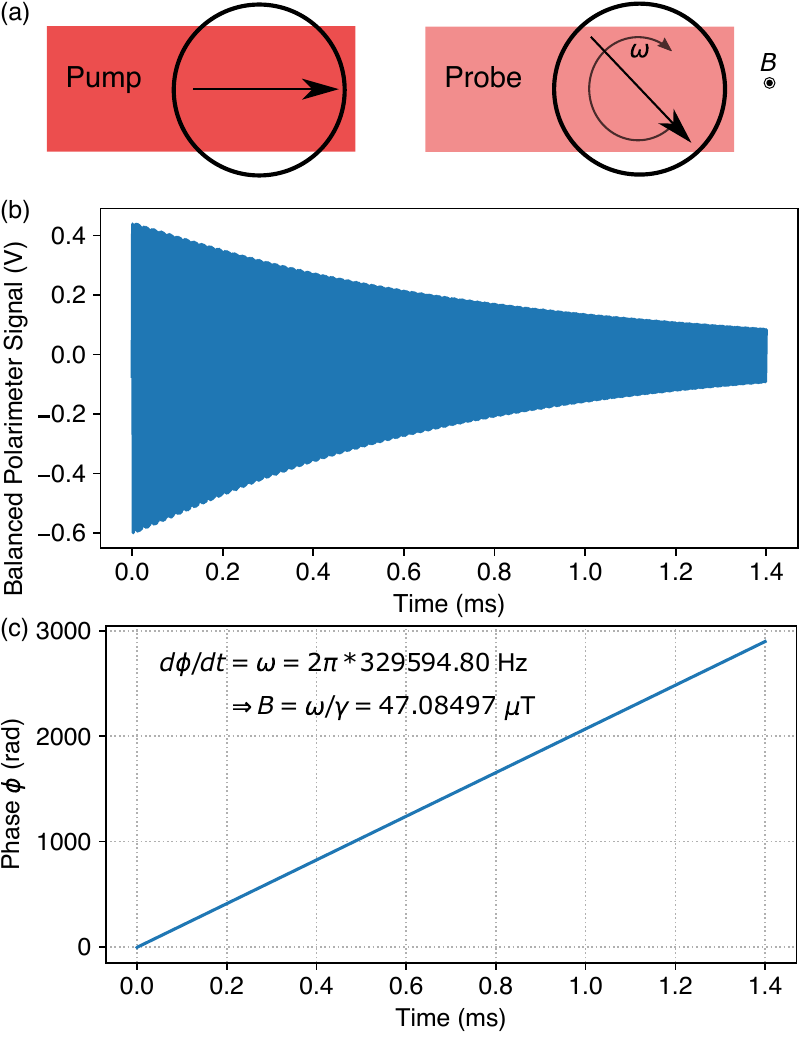}
\caption{ (a) The scalar $^{87}$Rb magnetometer is operated by pumping spins transverse to a bias field, then detecting the (b) free precession by a linearly polarized probe beam that undergoes Faraday rotation into a balanced polarimeter. (c) The free precession frequency $\omega$ over a repetition can be found by extracting the phase accumulation through a Hilbert transform of time-domain data, followed by a linear regression to the phase $\phi$ versus time $t$, to obtain the slope and thus frequency $\omega$ of the free precession decay.  This frequency can then be related to the magnetic field $B$ through the gyromagnetic ratio $\gamma$, $B = \omega /\gamma$. }
\label{fig:1}
\end{figure}


To form a scalar atomic free-precession magnetometer, $^{87}$Rb vapor is optically pumped with resonant 795 nm circularly polarized light from a non-magnetic, compact, narrowed edge-emitter semiconductor laser within the sensor head, to create large atomic polarizations ($>$80\%) in a plane transverse to a total magnetic field $B$ (see Fig.~\ref{fig:1}(a)). 
After a pumping period, a detuned linearly polarized probe beam from a VCSEL or DBR laser undergoes paramagnetic Faraday rotation due to its near-resonant interaction with the polarized $^{87}$Rb atoms in the transverse plane. 
The optical rotation of the light is detected using a balanced polarimeter consisting of a polarizing beamsplitter cube and two photodiodes that are each sent into transimpedance amplifiers and subtracted using an analog circuit. 
The resulting voltage time series data is then assumed to take the form of a exponentially decaying sine wave \cite{Limes_2020,Jaufenthaler_2021,Lee_2021, Lucivero_2022}, 
\begin{equation}
A \exp(t/T_2)\sin(\omega t +\phi_0).
\label{eq:decSine}
\end{equation}
While using Eq.~\ref{eq:decSine} is not entirely representative of the dynamics at Earth-scale fields of approximately 50 $\mu$T, due to the non-linear Zeeman splitting causing different precession frequency of the $F=2$ and $F=1$ manifolds, the $F=1$ is often small with quickly decaying signals relative to the $F=2$ state \cite{Lee_2021}, so the frequency chirp within the shot caused by the effect is neglected. Time-series data from a single cell of the analog output of a Twinleaf Optical Magnetic Gradiometer (OMG) is shown in Fig.~\ref{fig:1}(b).

Extracting a field is a task of finding the frequency in Eq.~\ref{eq:decSine} and dividing by the effective $^{87}$Rb gyromagnetic ratio. 
The signal is faithfully digitized if the dynamic range of the digitization can, for example, capture the maximum amplitude typically on the order of volts, while resolving the photon shot noise that may range from 20-200 of nV/Hz$^{1/2}$. If Gaussian photon shot noise is the dominant noise, the Cramer-Rao Lower Bound (CRLB) can be derived for the frequency error estimate $\sigma_f$ from fitting a decaying sine wave,
\begin{equation}
\sigma_f = \frac{\sqrt{12C}}{2\pi (A/\rho) T^{3/2}}~\text{Hz}, 
\label{eq:cramerRao}
\end{equation}
where $A$ is the initial sine wave amplitude, $\rho$ is the white noise in units of V/Hz$^{1/2}$, $T$ is the measurement time, and $C$ (in our limit of high sample rate) is a constant that takes into account the exponential decay relative to the measurement time $T$ \cite{Lucivero_2022}. 
By including measurement dead-time $T_d$, e.g.~from pumping and heating, we form a repetition rate $f_r = 1/T_r = 1/(T+T_d)$ to determine the bandwidth of the sensor, $\text{BW} = 1/2T_r$.
A bandwidth-free estimate of the scalar sensor sensitivity can then be obtained by taking $\sigma_f/\text{BW}^{1/2}$. 

Time series data may be fit to Eq.~\ref{eq:decSine} by using a non-linear least-squares regression scheme such as the Levenberg-Marquat algorithm, which interpolates between Gauss-Newton and gradient descent.  These schemes need to be seeded well, and can be computationally expensive, prohibiting a real-time portable fitting solution.

For real-time operation, pulsed free-precession sensors need fast fitting routines to extract a frequency from a decaying sine wave. 
One such method is to form the analytic signal from the voltage time series data $x$ by using a discrete Hilbert transform, or its approximate FIR filter \cite{Wilson_2020 , Jaufenthaler_2021}. 
Practically, the Hilbert transform amounts to taking a conventional fast Fourier transform, zeroing out the negative frequencies, and transforming back to the time-domain to obtain the analytic function $z =x + iy$. 
The accumulated phase $\phi$ can be obtained by taking $\phi = \arctan(y/x)$, with the same data rate as the digitization of the raw voltage signal.  
To obtain the precession frequency $\omega$ of the $^{87}$Rb, the slope $d\phi/dt = \omega$ can be found, and in turn the experienced magnetic field $B = \omega/\gamma$, where $\gamma/2\pi \approx 7$ GHz/T is the low-field gyromagnetic ratio of $^{87}$Rb, due to its spin-3/2 nuclear slowing down factor. 
The slope $d\phi/dt$ can be efficiently extracted using a linear regression, $\mathbf{y} = \mathbf{X}\beta$, where $\mathbf{y}$ is the phase data, $\beta$ contains the slope $m$ and intercept $b$ of the fit, and 
\begin{equation}
\mathbf{X}^T = 
\begin{bmatrix}
1 & 1 & \hdots & 1\\
t_0 & t_1 & \hdots & t_i\\
\end{bmatrix}.
\end{equation}
The regression can be solved by many different efficient methods such as an updating simple linear regression \cite{Klotz_1995}, or by using a precalculated matrix $\mathbf{\Omega} =(\mathbf{X}^T \mathbf{X} )^{-1}\mathbf{X}^T  $ to find $\beta =\mathbf{ \Omega}\mathbf{y} $.
Across many repetitions for the sensor, one then has a time series of magnetic field data with repetition rate $f_r$, where a spectral density can be found, and noise floor obtained by fitting to the power spectral density. 

\begin{figure}
\includegraphics[width=3.3in]{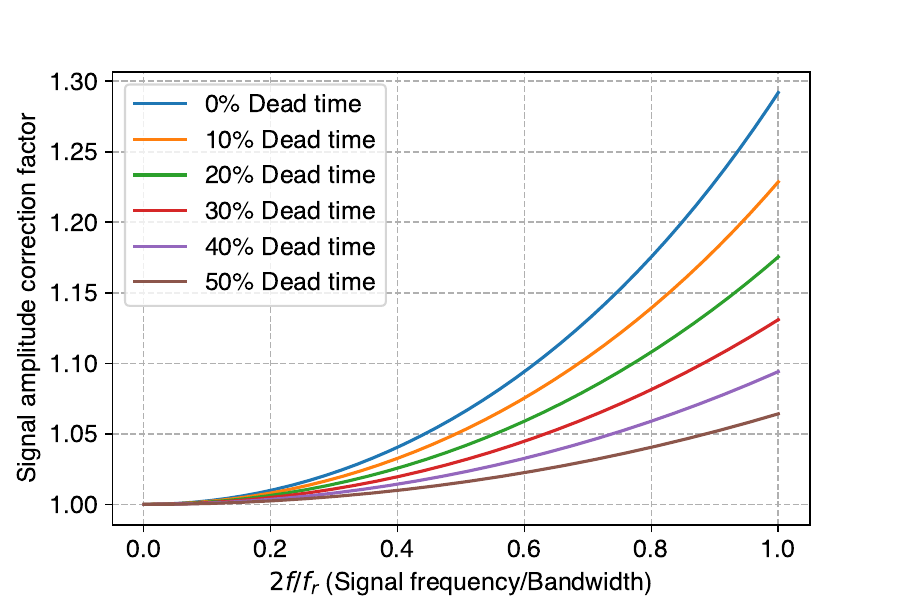}
\caption{The theoretical frequency dependent amplitude correction needed for the free-precession scalar magnetometer.  }
\label{fig:deadtimes}
\end{figure}

In previous works of pulsed free-precession sensors, it was assumed that this sensitivity was flat across the bandwidth of the sensor, i.e.,~no in-band, frequency-dependent degradation of response of the magnetometer. However, one must consider the effects of using a time-integrated measurement of data, as the instantaneous frequencies are averaged across a single repetition of the sensor, which leads to a decrease in perceived amplitude of sinusoidal waves being imparted on the sensor. (An analogous situation is that of reporting a $V_{\text{rms}}$ instead of a $V_{\text{peak}}$.)  
The results of any linear regression of time-dependent data to the form $\phi= mt + b$ can be expressed as   
\begin{equation}
m = \frac{\sum_{n = 0}^N t_i \phi_i}{\sum_{n = 0}^N t_i^2} = \frac{\overline{t\phi}}{\overline{t^2}},
\end{equation}
where, without loss of generality, we have offset the y-axis such that $b = 0$, and take the limit of the number of acquired data points to infinity and spacing between data points to zero. 
The phase accumulation of the $^{87}$Rb atoms can be expressed in terms of the magnetic field they experience by 
\begin{equation}
\phi = \gamma\int_{-T/2}^{T/2} B(t) dt, 
\label{eq:phase}
\end{equation}
where without loss of generality we have centered the x-axis about $t = 0$. 
Assuming static fields, we obtain the familiar expression $\phi = \gamma B t$, and the appropriate slope extracted from the linear regression of $m = \gamma B$. 
If we allow for oscillating fields along a $\hat{z}$ bias field $B$,we have $B(t) = B + b_z \sin(\omega_z t + \phi_z)$, resulting in a phase accumulation
\begin{equation}
\phi(t) = \gamma B t + \frac{\gamma b_z}{\omega_z}\sin(\omega_z t +\phi_z) .
\end{equation} 
Letting $\phi_z = 0$, we have the linear regression for the slope that now gives 
\begin{equation}
m = \gamma B + \gamma b_z \frac{12f_r^2}{\omega_z^2}(\text{sinc}\frac{\omega_z}{ 2 f_r } - \cos\frac{\omega}{ 2 f_r }) . 
\end{equation}
(Note a similar result is also obtained in Ref.~\cite{Rubiola_2016}.)
For the scalar magnetometer, one must also incorporate dead-time, with can be done with the substitution $\omega_z/2f_r \rightarrow \omega_z(1 - f_rT_d)/2f_r = \alpha$, resulting in an amplitude response of
\begin{equation}
\frac{3 }{\alpha^2}(\text{sinc}\alpha - \cos\alpha).
\label{eq:correction}
\end{equation}
In Fig.~\ref{fig:deadtimes} we show the effect of dead time on the signal amplitude correction. In the limit of 100\% dead time, the frequency measurement becomes instantaneous and there is no correction required. 
We note most free-precession magnetometers post their magnetic noise floors with logarithmic scaling along the sensitivity axis, making the effect hard to see in the spectral density analysis. 
The time-integrated field correction is needed for any fitting routine that assumes the model equation is a single frequency sine wave as in Eq.~\ref{eq:decSine}. 
There is no \emph{a priori} method to incorporate the magnetic field frequency dependence within such a fit, unless there is only a single frequency of interest, or a very narrow frequency band to be corrected. 
Thus, statistics need to be collected in order to generically correct time-domain data that contains many unknown frequencies at different phases and amplitudes, as per the time-frequency uncertainty relation.

To study the scalar correction, we again use an OMG, which uses two magnetic field measurements in two anodically bonded, microfabricated cells each with 3x3x3 mm$^3$ active volume and single optical axis. 
A single probe and single pump beam contained within the sensor head are combined, split, and sent into each cell, resulting in one axis as the sensor dead zone.
The OMG is controlled and processes data through a proprietary electronics board and control software. 
For frequency extraction, a similar linear regression is used as the above description, and can be programmed through a simple python script to change repetition rates and blanking times for the real-time fitting routine; altogether the magnetometers on the OMG have been shown to reach 0.1 pT/$\sqrt{\text{Hz}}$ with a bandwidth of 240 Hz. 

The OMG is placed within mu-metal shielding, along with a magnetic coil set that is able to provide up to about 60 $\mu$T along the z-axis. 
To generate sine waves, a Keysight 33210A AWG is programmed for a given frequency, resulting in a modulation along the bias field of about 14 nT in amplitude. 
Time-series magnetic field data are collected, and the magnetometer response to the modulation is fit to a sine wave using a non-linear fitting routine. 
\begin{figure}
\includegraphics[width=3.3in]{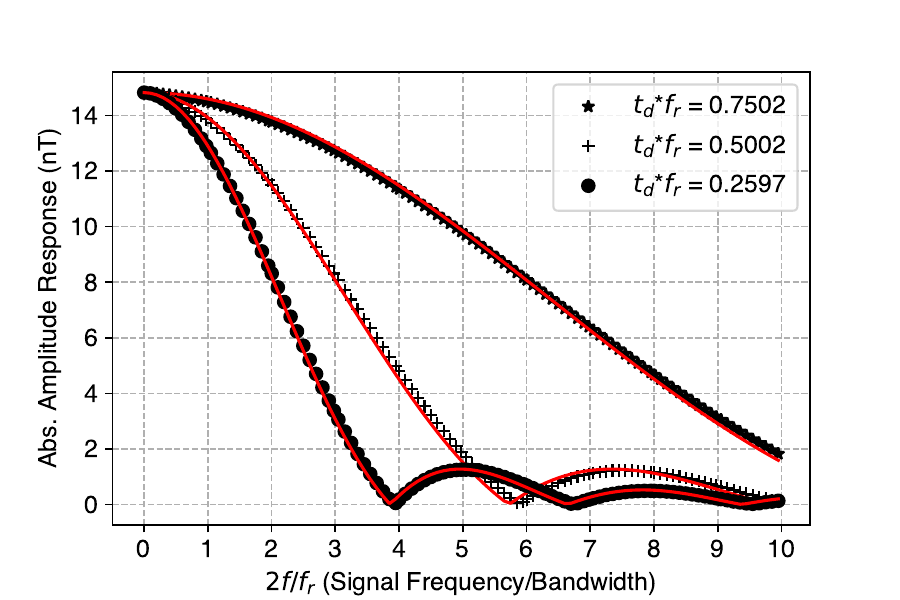}
\caption{Out-of-band AC waves are applied along the bias field axis of a free-precession sensor, with magnetic field time fit to sine waves to extract amplitudes. Three different dead time fractions are shown, along with the no-free-parameters theory of Eq.~\ref{eq:correction}.}
\label{fig:extendedBW}
\end{figure}
The results of a experimental run are shown in Fig.~\ref{fig:extendedBW}, where the response of the scalar magnetometer is shown outside of the bandwidth of the sensor, for three different dead-time fractions of roughly 25\%, 50\%, and 75\%. 
Here we see that, while the smallest dead-time has the greatest degradation of in-band response, it suffers the least from large aliasing of out-of-band signals. 
On the other hand, while there is no need for correction in the limit of an instantaneous frequency measurement when the dead-time approaches 100\%, there is also a very large aliasing of out-of-band signals. 
Plotted along with the experimental amplitudes in Fig.~\ref{fig:extendedBW} are the no-free-parameter curves of Eq.~\ref{eq:correction}. 
The zero crossings of Eq.~\ref{eq:correction} are sensitive to dead-time fraction, indicating a parameter-free fit within 1 \%; also at the zero-crossings there an response inversion, as shown in Ref.~\cite{Rubiola_2016}.

Here we propose a scheme to identify aliased frequencies by fitting the same magnetic field data stream with varying dead-time fractions. 
For example, assume the bandwidth of sensor is 1 kHz, and a 500 Hz signal is observed in the spectral density.
One can take the ratio of signal heights using different dead-time fractions to determine if the observed 500 Hz signal is instead a 1500 Hz or higher harmonic signal aliased in. 
A 1500 Hz signal will have a 75\% dead-time fit over 25\% dead-time fit amplitude ratio that is roughly a factor of about 1.4, compared to a 2500 Hz ratio of about 2.2. 
More than two dead-times can be used to avoid uniqueness issues, to create a method of fingerprinting high frequency aliased signals, and rejecting or utilizing them if needed. 
Theoretically, the loss of sensitivity by increasing the dead-time from 25\% to 75\% in Eq.~\ref{eq:cramerRao} is roughly a factor of 3-5, depending on the exact $T_2$ in relation to measurement time $T$. 
Using a fixed dead-time of 25\%, increasing the repetition rate in order to increase the bandwidth of the sensor results in a linear loss of sensitivity in proportional to the repetition rate. 
As another example, increasing the repetition rate from 2 kHz to 10 kHz, causes a loss of sensitivity of about a factor of 4-5, keeping all other system parameters the same (note that sensitivity can be recouped in the physical system by increasing the temperature of the cell to increase atom density helps while increasing the bandwidth, up to the limit of available pump power).  
Compare this to the situation of analyzing aliased data of a ~5 kHz signal using a 2 kHz repetition rate (1 kHz bandwidth) with 25\% dead-time, there is a loss in amplitude response of about 30\%, and as mentioned above, an inherent loss of sensitivity in the noise floor of about a factor of 3-5. 
One can also sample the 75\% dead-time data 3 times per shot, resulting in a $\sqrt{3}$ averaging advantage, leading to a overall loss in sensitivity of about 3-5. 
Thus, by identifying aliased signals, one can transform a lower bandwidth scalar sensor to a high-bandwidth sensor with similar results as increasing the repetition rate of the scalar sensor, while retaining higher sensitivity to low-frequency signals by retaining the low dead-time fit.

\begin{figure}
\includegraphics[width=3.3in]{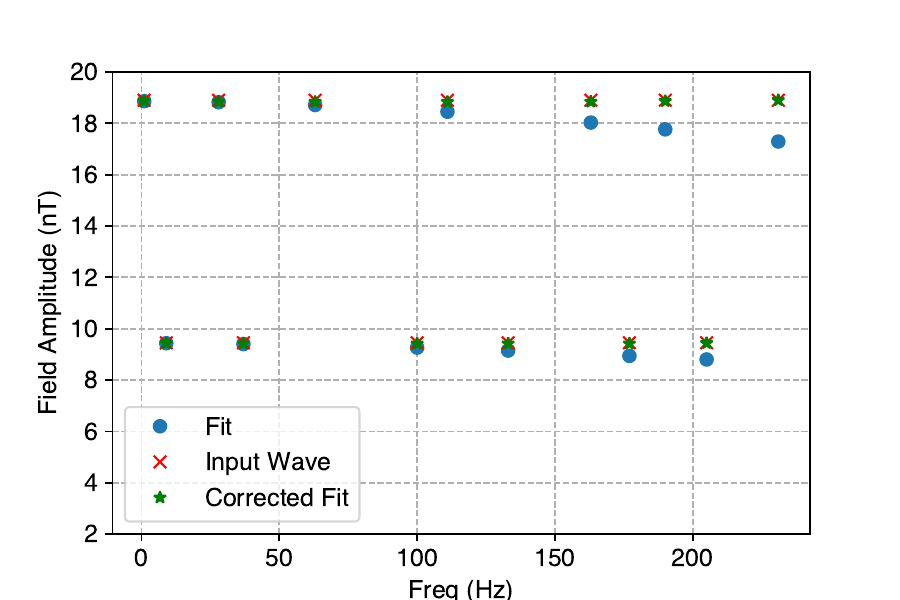}
\caption{Pulsed sensor data is taken with a sum of many sine waves of different phase, amplitude, and frequency. We show the raw magnetometer data can be generically corrected in the complex frequency domain and transformed back into the time-domain using Eq.~\ref{eq:corr}.}
\label{fig:CorrectedPeaks}
\end{figure} 
 
To correct the spectral density of the data, Eq.~\ref{eq:correction} can be applied in the complex frequency domain with desired frequency resolution as determined by the time-frequency uncertainty principle. For example, with the transform to $N$ time-series points where $\mathcal{F}$ can be a $N \times N$ matrix representing a discrete Fourier transform, and $\mathcal{C}$ is a set of $N$ dependent on $\omega_z$ where the constants of $\mathcal{C}$ are applied row-wise to $\mathcal{F}$. To obtain the corrected time domain data, an inverse transform can be applied, resulting in a transformation matrix $\Gamma$ of 
\begin{equation}
\Gamma = \mathcal{F}^{-1} \mathcal{C}\mathcal{F}. 
\label{eq:corr}
\end{equation} 
To test this methodology, the magnetic sensor is fed a sum of sine waves along the bias field, from a programmed AWG of different frequencies and phases. Here we used frequencies of $\{1, 9, 28, 37, 63, 100, 111, 133, 163, 177, 190, 205, 231\}$ Hz and phases \{0, 3/4, -3, 3/2, -3/8, 3/16, 0, 2, -3/4, -4/3, 4/3, 3/16, 3/32\}, with every other amplitude in the set being 18.9 and 9.43 nT amplitude. Data is recorded for 5 seconds, and then independent time-domain sine waves are fit to extract amplitudes. The correction $\Gamma$ is applied using a $t_d *f_r = 0.38$, and then the independent time-domain sine waves are again fit, and corrected amplitudes are extracted. Results are seen in Fig.~\ref{fig:CorrectedPeaks}, with the corrected amplitudes roughly matching the input wave amplitude. For real-time correction, it may be preferable to map this problem to an appropriate FIR filter representation.

We have shown that, due to the time-integrated measurement of fields used in spin-based free-precession magnetometers, there is a non-linear amplitude response across the magnetometer bandwidth. This correction can be applied to arbitrary phases and fields in the complex frequency domain. Moreover, this effect also provides a solution for aliasing out-of-band frequencies for such sensors. By correcting the frequency dependent sensor response, applications such as source localization in magnetoencephalography studies can be made more accurate. 

The authors thank M.~V.~Romalis for helpful discussions. This material is based upon work supported by the Defense Advanced Research Projects Agency (DARPA) under Contract N.~HR0011-20-C-0009. Any opinions, findings and conclusions or recommendations expressed in this material are those of the author(s) and do not necessarily reflect the views of DARPA.  Approved for Public Release, Distribution Unlimited.

%
%

\end{document}